\documentclass[sigconf]{acmart}
\usepackage{enumitem}
\usepackage{makecell}
\usepackage{float}

\AtBeginDocument{%
  }

\copyrightyear{2025}
\acmYear{2025}
\setcopyright{cc}
\setcctype{by}
\acmConference[UbiComp Companion '25]{Companion of the the 2025 ACM International Joint Conference on Pervasive and Ubiquitous Computing}{October 12--16, 2025}{Espoo, Finland}
\acmBooktitle{Companion of the 2025 ACM International Joint Conference on Pervasive and Ubiquitous Computing (UbiComp Companion '25), October 12--16, 2025, Espoo, Finland}\acmDOI{10.1145/3714394.3756254}
\acmISBN{979-8-4007-1477-1/2025/10}

\begin{document}

%%
%% The "title" command has an optional parameter,
%% allowing the author to define a "short title" to be used in page headers.
\title{WHAR Datasets: An Open Source Library for \\ Wearable Human Activity Recognition}

%%
%% The "author" command and its associated commands are used to define
%% the authors and their affiliations.
%% Of note is the shared affiliation of the first two authors, and the
%% "authornote" and "authornotemark" commands
%% used to denote shared contribution to the research.
\author{Maximilian Burzer}
% \authornote{Both authors contributed equally to this research.}
\email{maximilian.burzer@kit.edu}
\affiliation{%
  \institution{Karlsruhe Institute of Technology}
  \city{Karlsruhe}
  \country{Germany}
}

\author{Tobias King}
% \authornote{Both authors contributed equally to this research.}
\email{tobias.king@kit.edu}
\affiliation{%
  \institution{Karlsruhe Institute of Technology}
  \city{Karlsruhe}
  \country{Germany}
}

\author{Till Riedel}
% \authornote{Both authors contributed equally to this research.}
\email{riedel@kit.edu}
\affiliation{%
  \institution{Karlsruhe Institute of Technology}
  \city{Karlsruhe}
  \country{Germany}
}

\author{Michael Beigl}
% \authornote{Both authors contributed equally to this research.}
\email{michael.beigl@kit.edu}
\affiliation{%
  \institution{Karlsruhe Institute of Technology}
  \city{Karlsruhe}
  \country{Germany}
}

\author{Tobias Röddiger}
% \authornote{Both authors contributed equally to this research.}
\email{tobias.roeddiger@kit.edu}
\affiliation{%
  \institution{Karlsruhe Institute of Technology}
  \city{Karlsruhe}
  \country{Germany}
}

%%
%% By default, the full list of authors will be used in the page
%% headers. Often, this list is too long, and will overlap
%% other information printed in the page headers. This command allows
%% the author to define a more concise list
%% of authors' names for this purpose.
\renewcommand{\shortauthors}{Maximilian Burzer, Tobias King, Till Riedel, Michael Beigl, \& Tobias R\"{o}ddiger}

%%
%% The abstract is a short summary of the work to be presented in the
%% article.
\begin{abstract}
The lack of standardization across Wearable Human Activity Recognition (WHAR) datasets limits reproducibility, comparability, and research efficiency. We introduce WHAR datasets, an open-source library designed to simplify WHAR data handling through a standardized data format and a configuration-driven design, enabling reproducible and computationally efficient workflows with minimal manual intervention. The library currently supports 9 widely-used datasets, integrates with PyTorch and TensorFlow, and is easily extensible to new datasets. To demonstrate its utility, we trained two state-of-the-art models, TinyHar and MLP-HAR, on the included datasets, approximately reproducing published results and validating the library’s effectiveness for experimentation and benchmarking. Additionally, we evaluated preprocessing performance and observed speedups of up to 3.8× using multiprocessing. We hope this library contributes to more efficient, reproducible, and comparable WHAR research.
\end{abstract}

%%
%% The code below is generated by the tool at http://dl.acm.org/ccs.cfm.
%% Please copy and paste the code instead of the example below.
%%


%%
%% Keywords. The author(s) should pick words that accurately describe
%% the work being presented. Separate the keywords with commas.
%% A "teaser" image appears between the author and affiliation
%% information and the body of the document, and typically spans the
%% page.
% \begin{teaserfigure}
%   \includegraphics[width=\textwidth]{sampleteaser}
%   \caption{Seattle Mariners at Spring Training, 2010.}
%   \Description{Enjoying the baseball game from the third-base
%   seats. Ichiro Suzuki preparing to bat.}
%   \label{fig:teaser}
% \end{teaserfigure}

\received{17 July 2025}
% \received[revised]{12 March 2009}
\received[accepted]{21 July 2025}

%%
%% This command processes the author and affiliation and title
%% information and builds the first part of the formatted document.
\maketitle

\section{Introduction}

% Background & Problem
Wearable Human Activity Recognition (WHAR) research in areas such as healthcare, sports analytics, and smart environments \cite{huang_survey_2024} has led to the creation of a diverse and growing number of datasets \cite{uci-har, wisdm, opportunity}. However, this growth has also exposed a significant challenge within the WHAR community: the lack of standardization. Datasets are often characterized by significant variability in file structures and formats, heterogeneous data, and general preprocessing requirements \cite{alam_open_2023, wisdm, uci-har, opportunity}. As a result, researchers frequently rely on custom dataset-specific code to handle and experiment with each dataset \cite{alam_open_2023}. This leads to repetitive work, hinders the reproducibility of published results, and complicates fair comparisons across models, even when evaluated on the same dataset. Unlike other fields such as natural language processing, where centralized benchmarks are common, the lack of standardization in WHAR makes such efforts difficult. Consequently, researchers still spend considerable time and effort on the tedious and error-prone task of data handling, diverting focus from their primary research goals and novel contributions. Although the need for standardization has been acknowledged in the literature \cite{alam_open_2023, huang_standardizing_2024}, only limited steps have been taken to tackle these issues systematically.

% Solution & Contributions
To overcome these challenges, we introduce a novel open-source library designed to standardize and streamline data handling in WHAR research. The library is available on GitHub under the MIT license \cite{whar_datasets_lib}, encouraging community collaboration and wide adoption. Our library allows the conversion of a growing number of datasets from the literature into a standardized format using dataset-specific parsers, which can subsequently be preprocessed and loaded without requiring manual work. Using multiprocessing and caching, the library achieves both computational and memory efficiency, minimizing redundant recomputation. It is framework-agnostic at its core, to allow integration with different deep learning frameworks such as PyTorch \cite{pytorch} and TensorFlow \cite{tensorflow2015-whitepaper} using framework-specific adapters. A core strength of our library is its configuration-driven design, enabling consistent and unified data handling while remaining flexible to the unique requirements of each dataset. We designed the library in such a way that custom WHAR datasets, which are not yet supported, can easily be integrated and used with the library by simply providing a dataset-specific configuration including a parser. By abstracting away the complexities and inconsistencies of individual dataset formats, this approach aims to provide a unified data handling solution for WHAR. We hope that the introduction of this standardized library will lead to the following advancements within the WHAR research community:

\begin{enumerate}
    \item \textbf{Increased Efficiency}: By providing ready-to-use data handling, the library aims to allow researchers to focus on building and evaluating models instead of spending time on data handling and integration.
    \item \textbf{Improved Reproducibility}: By enforcing consistent data configurations, our library is expected to help researchers replicate experiments more reliably, reducing the variability introduced by inconsistent data handling.
    \item \textbf{Fair Comparability}: Standardized data handling has the potential to enable fair and direct comparisons across models and datasets, supporting the development of standardized WHAR benchmarks.
\end{enumerate}

% \item \textbf{Standardization}: The library is intended to encourage dataset authors to adopt a standardized data format when releasing new WHAR datasets, reducing the integration effort and fostering a more cohesive research ecosystem.
\section{Background and Motivation}

WHAR relies heavily on the availability and reuse of open datasets \cite{alam_open_2023}. However, numerous challenges hinder effective utilization, as identified by \cite{alam_open_2023} through a comprehensive literature review and a questionnaire survey of researchers with expertise in HAR. Their study presents a conceptual framework encompassing four phases in the open dataset lifecycle: construction, sharing, searching, and usage. Across these phases, several recurring issues were identified, that significantly impact research productivity and reproducibility.

One of the most pressing challenges highlighted is the absence of a standardized data format for HAR datasets. This lack of consistency complicates data sharing, hinders reproducibility, and adds unnecessary effort when benchmarking or applying novel methods. Researchers reported that datasets often come with missing values, errors, or in unstructured, non-presentable formats, discouraging reuse and increasing the preprocessing burden. In fact, when using existing datasets, nearly 67.7\% of participants indicated needing either “some” or “a lot” of preprocessing effort before the data could be used experimentally. Moreover, while a large majority (96.9\%) of respondents download open datasets for experimentation, their ability to make use of them is limited by poor metadata, inconsistent annotations, and idiosyncratic formats. Key selection criteria for dataset reuse include the availability of code or scripts for data processing (48.4\%), the data format itself (54.8\%), and the presence of clear metadata (45.2\%). These concerns highlight the community’s desire for datasets that are easy to integrate, well-documented, and usable with minimal manual intervention.

Our work directly addresses these gaps by introducing a novel open-source library designed to standardize and streamline dataset handling in WHAR research. It tackles the root causes behind many of the data-sharing and usability issues, such as lack of standardization, metadata inconsistency, and preprocessing overhead, ultimately promoting collaboration and reproducibility in the WHAR community.

\section{Related Work}

Efforts to standardize dataset access and improve usability have gained significant traction across machine learning domains. Prominent general-purpose libraries and platforms such as Hugging Face \cite{lhoest-etal-2021-datasets} and OpenML \cite{JMLR:v22:19-920} have made substantial contributions in this area. Hugging Face’s Datasets library enables seamless loading of datasets for natural language processing, computer vision, and audio tasks via a unified API. Integration with the Hugging Face Hub further simplifies sharing and collaboration within the community. Similarly, OpenML serves as an open platform for sharing datasets, provides standardized APIs, and encourages reproducibility and reuse through its collaborative design. While both platforms promote dataset accessibility and interoperability, their primary focus is on general-purpose usage across domains. They neither offer domain-specific preprocessing pipelines nor support the specialized requirements of WHAR datasets, e.g. resampling or windowing. 

Several domain-specific libraries address adjacent challenges. For instance, \cite{herzen2022darts} and sktime \cite{löning2019sktimeunifiedinterfacemachine} offer comprehensive frameworks for time series learning, supporting forecasting, classification, and transformation pipelines. While including preprocessing functionality relevant to WHAR, their focus remains on general time series tasks. As such, these frameworks fall short in addressing the domain-specific challenges of WHAR pipelines and cannot be directly applied without extensive adaptation. 

Other initiatives such as MLCroissant \cite{Akhtar_2024} and DCAT-AP \cite{dcat-ap} focus on standardizing dataset metadata and improving dataset discoverability, particularly in the context of FAIR (findability, accessibility, interoperability, reusability) data principles \cite{wilkinson2016fair}. While these formats contribute to making datasets more “AI-ready,” they are primarily concerned with semantic interoperability rather than the structural readiness required for WHAR data handling.

The lack of standardization in WHAR has also been addressed by prior work such as \cite{huang_standardizing_2024}, which focuses primarily on standardizing training methodologies and evaluation protocols. However, their contribution centers on the modeling and training process rather than on dataset access and preprocessing.

In contrast to these prior efforts, our proposed library is purpose-built for WHAR data handling. It unifies dataset access, preprocessing, and deep learning integration through a modular architecture centered around a standardized data format. Crucially, our focus on WHAR-specific requirements distinguishes our library from general-purpose solutions. Over time, this framework can serve as a foundation for benchmarking and sharing WHAR datasets in a consistent and reproducible manner.

\section{Requirements and Features}

To help achieve the goals of increased efficiency, improved reproducibility, fair comparability, the WHAR datasets library must satisfy several key design requirements and offer a rich set of features. We begin by outlining the non-functional requirements.

\begin{enumerate}[label=(R\arabic*)]
    \item \textbf{Usability}: The library should offer a simple, intuitive interface for preprocessing and loading WHAR datasets, enabling users to start experiments with just a few lines of code.
    \item \textbf{Reproducibility}: A standardized, dataset-agnostic configuration schema should define data handling parameters transparently, ensuring consistent and reusable experimentation.
    \item \textbf{Dataset-Agnosticism}: The library should support diverse WHAR datasets by converting them into a standardized format. Integrating new datasets should require minimal effort, using a parser to map to this format, ensuring extensibility and backward compatibility.
    \item \textbf{Framework-Agnosticism}: The library should be compatible with major deep learning frameworks like PyTorch~\cite{pytorch} and TensorFlow~\cite{tensorflow2015-whitepaper}, allowing flexible model development.
    \item \textbf{Computational Efficiency}: To support large datasets and changing hyperparameters, the standardized data format should be optimized for computational efficiency, enabling multiprocessing and caching to ensure scalability.
\end{enumerate}

In addition to these non-functional requirements, the following functional features are critical. They are tailored specifically to the WHAR domain and underscore the necessity of a dedicated library designed to address its unique challenges.

\begin{enumerate}[label=(F\arabic*)]
    \item \textbf{Subject-Wise Splitting}: The library must support subject-disjoint train/validation/test splits to prevent data leakage and better reflect real-world deployment. It should enable protocols like subject-wise and leave-one-subject-out (LOSO) cross-validation.
    \item \textbf{Normalization}: Built-in support for common normalization methods, including min-max, z-score, and robust scaling, is required. Both per-window and global (train-set-based) normalization modes should be supported.
    \item \textbf{Sample Loading}: To accommodate varying memory and runtime constraints, the library should provide two sample-loading strategies: on-demand (loading data from disk as needed) for large datasets, and preload (loading all windows into memory) for faster access when dealing with smaller datasets, which is common for WHAR.
    \item \textbf{Class Weighting}: Given the prevalence of class imbalance in WHAR due to varying activity durations, the library should compute class weights from the training data for integration into loss functions.
\end{enumerate}

\section{Design and Implementation}

\begin{figure}[ht]
  \centering
  \includegraphics[width=\linewidth]{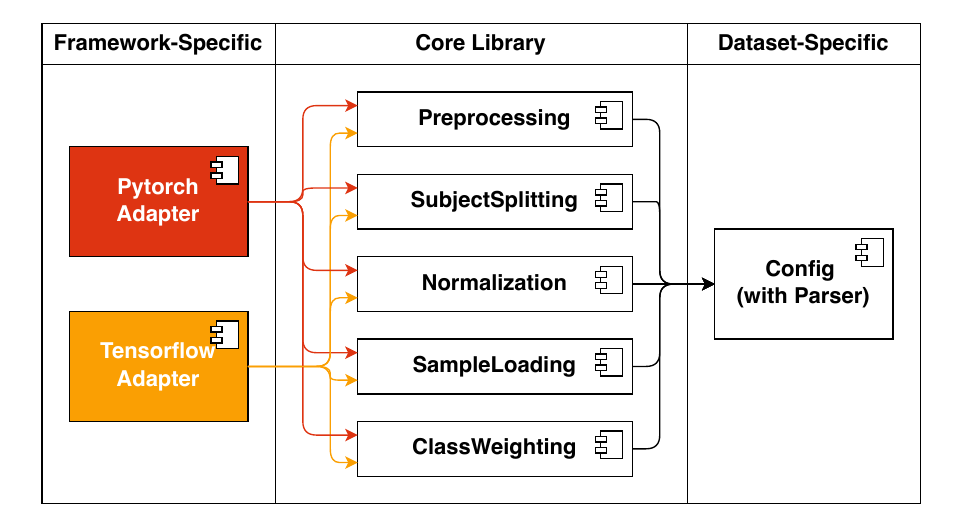}
  \caption{Structure of the library showing clear separation of dataset- and framework-specific components from feature components of the core library.}
  \label{fig:library-structure}
  \Description{
This diagram shows the modular architecture of a machine learning data preprocessing library.

The structure is organized into three main sections:

1. Core Library:
   - Modules:
     - SubjectSplitting
     - SampleLoading
     - Normalization
     - ClassWeighting
     - Preprocessing
   - These modules represent reusable, model-agnostic data processing components.

2. Framework-Specific Adapters:
   - TensorFlow Adapter
   - PyTorch Adapter
   - These components interface the core library with specific machine learning frameworks.

3. Dataset-Specific Configuration:
   - A Config component with an associated Parser.
   - This part provides customization and parsing logic tailored to specific datasets.

Arrows indicate directional dependencies:
- Dataset-specific configuration flows into the Core Library.
- The Core Library connects to framework-specific adapters, allowing flexible integration.
}
\end{figure}

Building on the non-functional requirements and functional features outlined in the previous section, the WHAR datasets library is designed with a strong focus on modularity, extensibility, and computational efficiency. As illustrated in Figure~\ref{fig:library-structure}, the architecture cleanly separates dataset-specific configuration and parsing, as well as framework-specific adapters, from the core library. This separation enables straightforward integration of new WHAR datasets (R3), ensures compatibility with major deep learning frameworks (R4), thereby enhancing overall usability (R1), and offers robust support for WHAR-specific data handling needs (F1–F4).

\subsection{Standardized Data Format}

To support scalable integration of diverse WHAR datasets (R3), the library employs a standardized data format based on a session-centric representation. Each session corresponds to a single subject performing a single activity and is stored as an individual file containing timestamp-indexed, multivariate time series data from inertial measurement units (IMUs) and other sensors. This design enables multiprocessing of sessions, enhancing the library’s scalability to very large datasets (R5). 

Parquet was selected as the storage format due to its columnar structure, built-in compression, and strong compatibility with scalable data processing frameworks such as Dask~\cite{rocklin2015dask}, all of which enhance computational efficiency and scalability (R5). Its suitability for large-scale time series machine learning tasks has also been demonstrated in a prior evaluation by OpenML~\cite{ openml_dataset_format_2020}. 

\begin{figure}[ht]
  \centering
  \includegraphics[width=\linewidth]{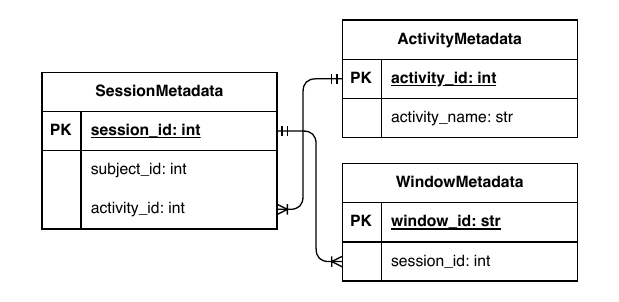}
  \caption{Entity-relationship diagram illustrating the metadata schema for the standardized data format.}
  \label{fig:erd-schema}
  \Description{
This Entity-Relationship Diagram (ERD) illustrates the relationships between three entities involved in modeling user activity and window interactions.

1. ActivityMetadata:
   - Attributes:
     - activity\_id (Primary Key, integer)
     - activity\_name (string)

2. WindowMetadata:
   - Attributes:
     - window\_id (Primary Key, string)
     - session\_id (integer)

3. SessionMetadata:
   - Attributes:
     - session\_id (Primary Key, integer)
     - subject\_id (integer)
     - activity\_id (integer)

Relationships:
- session\_id in WindowMetadata is a foreign key referencing session\_id in SessionMetadata.
- activity\_id in SessionMetadata is a foreign key referencing activity\_id in ActivityMetadata.

The layout is hierarchical, with ActivityMetadata and WindowMetadata at the top and SessionMetadata in the middle, serving as a link between the two.
}
\end{figure}

To support features such as subject-wise data splitting (F1) and class weighting (F4) without repeated reads of raw sensor data, metadata required for subject identification and activity labeling is stored separately in centralized, structured tables. This metadata layer, illustrated in Figure~\ref{fig:erd-schema}, enables efficient filtering and partitioning during preprocessing, further contributing to overall computational efficiency (R5).

\begin{figure*}[ht]
  \centering
  \includegraphics[width=0.9\linewidth]{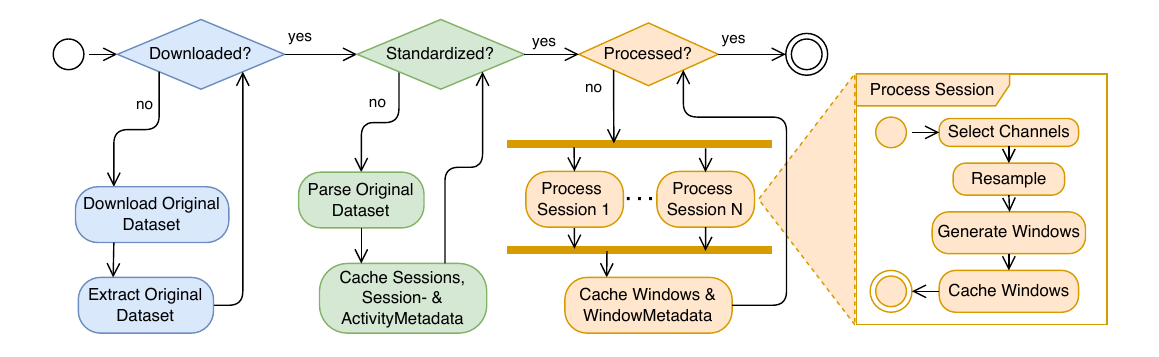}
  \caption{Activity diagram illustrating the preprocessing workflow. The process is divided into three main steps: (1) downloading the dataset, (2) converting it into a standardized data format, and (3) processing sessions to produce windows, optionally using multiprocessing. Each step is cached to prevent redundant computations.}
  \label{fig:preprocessing}
  \Description{
This flowchart illustrates the refactored data processing pipeline for handling time-series datasets in a machine learning context.

The pipeline begins with a decision point: "Downloaded?" with options "no" and "yes".  
- If "no", the pipeline initiates "Download Original Dataset".  
- If "yes", it checks: "Processed?"  
  - If "no", it proceeds to:
    - "Resample"
    - "Select Channels"
    - "Generate Windows"
    - Then, the flow leads to "Cache Windows & WindowMetadata"
  - If "yes", it checks: "Standardized?"  
    - If "no", the flow continues with:
      - "Resample"
      - "Select Channels"
    - If "yes", it proceeds to "Generate Windows" and then to "Cache Windows"

From "Download Original Dataset", an alternate branch flows to:
- "Parse Original Dataset"
- Followed by multiple session-specific processing steps:
  - "Process Session 1"
  - "Process Session"
  - "Process Session N"

These feed into:
- "Cache Sessions, Session- & ActivityMetadata"
- "Extract Original Dataset"

The layout reflects both decision-based branching and sequential processing stages, combining condition checks, transformation steps, and caching for reuse.
}
\end{figure*}

\subsection{Configuration and Parsing}

WHAR datasets are integrated through lightweight configurations containing dataset-specific metadata, data handling hyperparameters, and a parser. While implementing a parser requires an initial manual effort, it guarantees dataset-agnosticism (R3) by converting raw dataset formats into the standardized data format described earlier. Once integrated, the dataset is fully compatible with the core library’s preprocessing pipeline and feature components discussed in the next section. Users can easily adapt preprocessing by modifying configuration values, thereby supporting usability (R1). Moreover, using the same configuration with the library ensures consistent results, which promotes reproducibility (R2). 

\begin{figure}[!ht]
  \centering
  \includegraphics[width=0.98\linewidth]{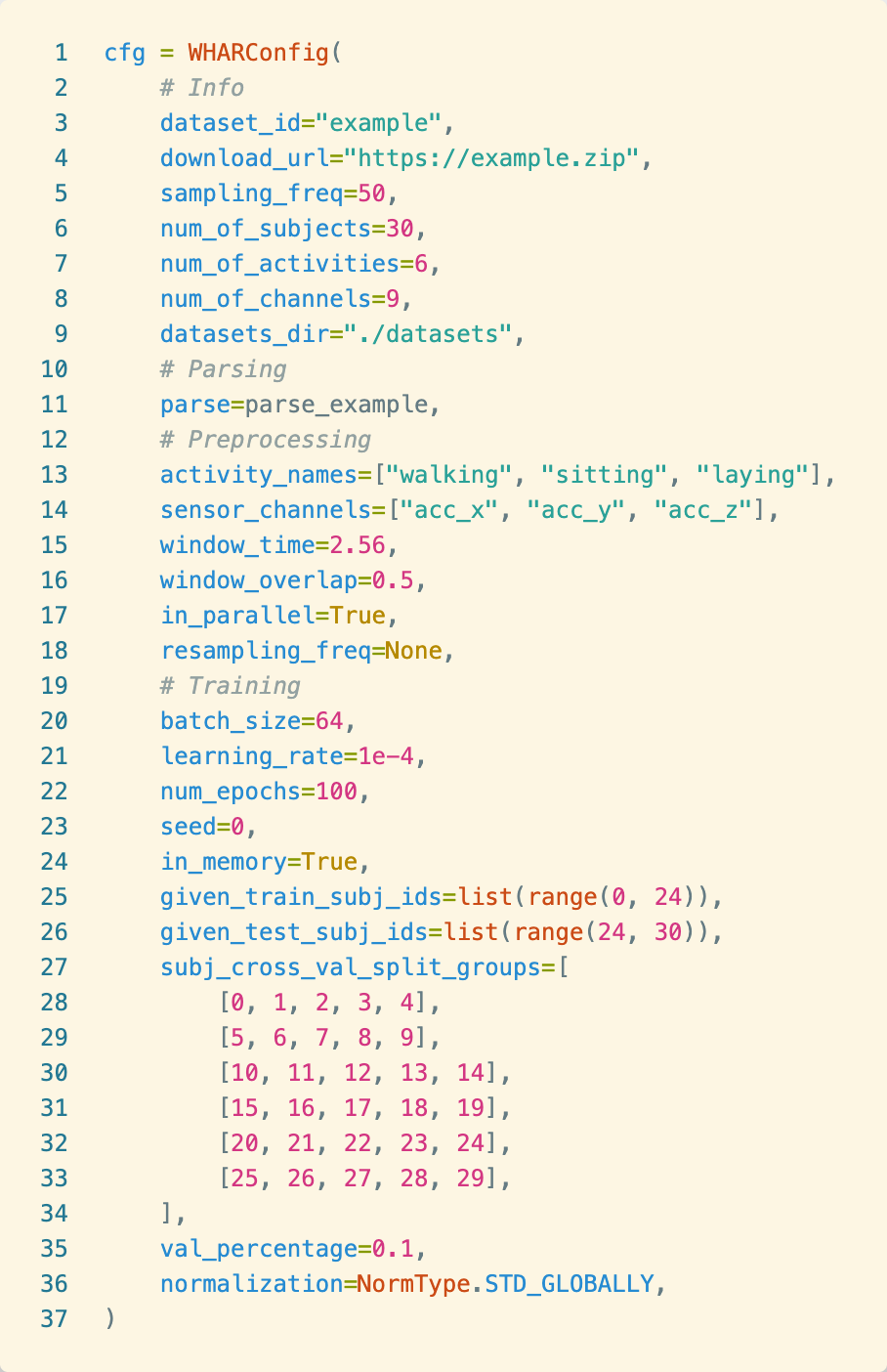}
  \caption{An example configuration containing metadata and hyperparameters, organized into sections for information, parsing, preprocessing, and training.}
  \label{fig:config}
  \Description{This figure displays a Python configuration file, likely for a Wearable Human Activity Recognition (WHAR) project, named \texttt{config.png}. The configuration, defined within a \texttt{WHARConfig()} object, is structured into several sections:
  \begin{itemize}
    \item \textbf{Info:} Specifies \texttt{dataset\_id} as "example", provides a \texttt{download\_url} for "example.zip", sets \texttt{sampling\_freq} to 50, \texttt{num\_of\_subjects} to 30, \texttt{num\_of\_activities} to 6, \texttt{num\_of\_channels} to 9, and defines \texttt{datasets\_dir} as "./datasets".
    \item \textbf{Parsing:} Includes a \texttt{parse} function, \texttt{parse\_example}.
    \item \textbf{Preprocessing:} Lists \texttt{activity\_names} as "walking", "sitting", and "laying", \texttt{sensor\_channels} as "acc\_x", "acc\_y", and "acc\_z", sets \texttt{window\_time} to 2.56, \texttt{window\_overlap} to 0.5, \texttt{in\_parallel} to True, and \texttt{resampling\_freq} to None.
    \item \textbf{Training:} Configures \texttt{batch\_size} to 64, \texttt{learning\_rate} to 1e-4, \texttt{num\_epochs} to 100, and \texttt{seed} to 0. It also specifies \texttt{in\_memory} as True, \texttt{given\_train\_subj\_ids} as a list from 0 to 23, \texttt{given\_test\_subj\_ids} as a list from 24 to 29, and \texttt{subj\_cross\_val\_split\_groups} which divides the 30 subjects into six groups of five subjects each for cross-validation.
  \end{itemize}
  Finally, \texttt{val\_percentage} is set to 0.1, and \texttt{normalization} is \texttt{NormType.STD\_GLOBALLY}.}
\end{figure}

To further improve usability and ensure configuration correctness (R1), the configuration schema is implemented using Pydantic, enabling automatic validation. This allows users to catch misconfigurations early and enhances the overall reliability of data handling. An example configuration is shown in Figure~\ref{fig:config}.

\subsection{Feature Components}

The core library implements a modular preprocessing pipeline alongside components that realize various functional features. These components operate on the standardized data format and are fully driven by dataset-specific configurations described in the previous section, promoting reproducibility across experiments (R2). Functional features, including subject-wise splitting (F1), normalization (F2), sample loading (F3), and class weighting (F4), are implemented as individual components. This design ensures consistent behavior across datasets while effectively addressing WHAR-specific data processing requirements.

The preprocessing pipeline handles the complete dataset preparation workflow, from downloading and parsing into the standardized format to executing a series of user-defined operations such as activity filtering, channel selection, resampling, and windowing. It validates the output of parsing against constraints of the standardized data format, utilizes caching to prevent redundant processing, and employs optional multiprocessing, ensuring efficiency and scalability (R4). The resulting data windows are saved as individual Parquet files, each linked to its session via a centralized metadata table (see Figure~\ref{fig:erd-schema}), ensuring full data traceability. An overview of this pipeline is shown in Figure~\ref{fig:preprocessing}.

\subsection{Framework Integration}

\begin{figure}[!ht]
\centering
\includegraphics[width=\linewidth]{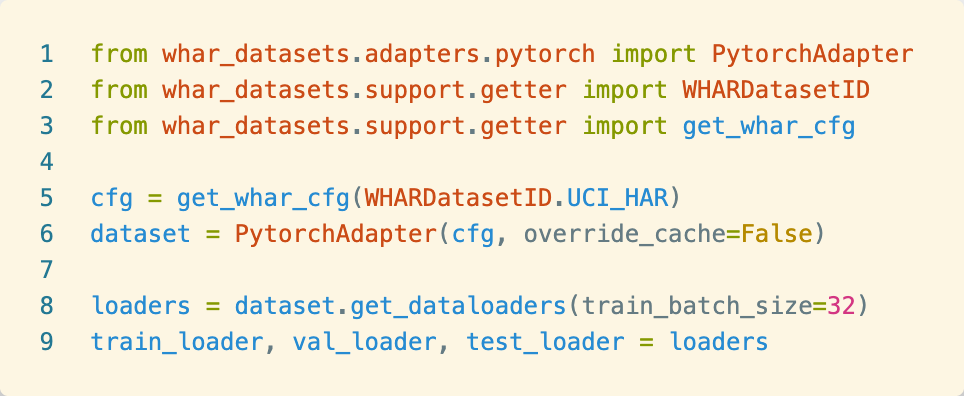}
\caption{Example usage of the library with PyTorch.}
\label{fig:pytorch-example}
\Description{This figure displays a Python code snippet demonstrating how to load and prepare a dataset for use with PyTorch, likely within the context of a Wearable Human Activity Recognition (WHAR) project.
  \begin{itemize}
    \item Line 1 imports \texttt{PytorchAdapter} from \texttt{whar\_datasets.adapters.pytorch}.
    \item Line 2 imports \texttt{WHARDatasetID} from \texttt{whar\_datasets.support.getter}.
    \item Line 3 imports \texttt{get\_whar\_cfg} from \texttt{whar\_datasets.support.getter}.
    \item Line 5 initializes a configuration object \texttt{cfg} by calling \texttt{get\_whar\_cfg} with \texttt{WHARDatasetID.UCI\_HAR}, indicating the use of the UCI HAR dataset.
    \item Line 6 creates a \texttt{PytorchAdapter} instance named \texttt{dataset}, passing the configuration \texttt{cfg} and setting \texttt{override\_cache} to False.
    \item Line 8 retrieves data loaders from the \texttt{dataset} object using \texttt{get\_dataloaders} with a \texttt{train\_batch\_size} of 32.
    \item Line 9 unpacks the returned loaders into three variables: \texttt{train\_loader}, \texttt{val\_loader}, and \texttt{test\_loader}, suggesting the creation of training, validation, and testing data loaders, respectively.
  \end{itemize}}
\end{figure}

To meet the requirement of framework-agnosticism (R4), the library provides dedicated adapters for the popular deep learning frameworks PyTorch~\cite{pytorch} and TensorFlow~\cite{tensorflow2015-whitepaper}. These adapters utilize the core library’s components while presenting familiar, framework-specific interfaces, enabling easy integration with existing training workflows and thus enhancing usability (R1). As illustrated in Figure~\ref{fig:pytorch-example}, all core library functionality is fully abstracted and driven by configuration, ensuring consistency and reproducibility across experiments.

\subsection{Supported Datasets}

To ensure immediate usability, the library includes built-in support for a set of WHAR datasets commonly used in the literature. These datasets vary widely in the number of subjects, activities, and sampling rates, as summarized in Table~\ref{tab:whar-datasets}. Each dataset comes with a complete configuration and a dedicated parser, enabling out-of-the-box use and enhancing usability (R1). These implementations also serve as practical templates for users who wish to integrate additional datasets. Note that our current focus has been on integrating a broad range of datasets by converting them into the standardized format. While we have applied basic data cleaning, we intend to enhance these efforts by refining and updating the parsers in the future.

\begin{table}[htb]
\centering
\caption{WHAR datasets currently supported by the library.}
\begin{tabular}{c|c|c|c}
\hline
\makecell{WHAR \\ Dataset} & \makecell{Number of \\ Subjects} & \makecell{Number of \\ Activities} & \makecell{Sampling \\ Rate (Hz)} \\ \hline
UCI-HAR~\cite{uci-har} & 30 & 6 & 50  \\ 
WISDM~\cite{wisdm} & 36 & 6 & 20  \\
MHEALTH~\cite{mhealth} & 10 & 12 & 50  \\
PAMAP2~\cite{pamap2} & 9 & 18 & 100  \\
OPPORTUNITY~\cite{opportunity} & 4 & 3 & 30  \\
MotionSense~\cite{motionsense} & 9 & 6 & 50  \\
DSADS~\cite{dsads} & 8 & 19 & 25  \\
Daphnet~\cite{daphnet} & 10 & 2 & 64  \\
% KU-HAR~\cite{ku-har} & 90 & 18 & 100  \\
HARSense~\cite{harsense} & 12 & 6 & 25  \\ \hline
\end{tabular}
\label{tab:whar-datasets}
\end{table}

Thanks to the library’s modular architecture, extending support to new datasets generally requires only implementing a parser and specifying a configuration, without modifying the core codebase. Potential datasets that could be integrated by the community to expand this collection include SHL~\cite{Wang2019shl}, RealLifeHar~\cite{GarciaGonzalez2020reallifehar}, ExtraSensory~\cite{Vaizman2017extrasensory}, RealWorld~\cite{Sztyler2016realworld}, UTD-MHAD~\cite{Chen2015utdmhad}, USC-HAD~\cite{Zhang2012uschad}, HuGaDB~\cite{Chereshnev2018hugadb}, w-HAR~\cite{Bhat2020whar}, HAPT~\cite{Reyes2016hapt}, and WISDM-19~\cite{Reyes-Ortiz2016wisdm}.

% \begin{table*}[htb]
% \centering
% \caption{Summary of WHAR datasets natively supported by the library.}
% \begin{tabular}{c|c|c|c|c}
% \hline
% WHAR Dataset & Subjects & Activities & Sensors & Sampling Rate (Hz) \\ \hline
% UCI-HAR~\cite{uci-har} & 30 & 6 & accel, gyro & 50  \\ 
% WISDM~\cite{wisdm} & 36 & 6 & accel & 20  \\
% MHEALTH~\cite{mhealth1, mhealth2} & 10 & 12 & 3x accel, 2x gyro, 2x mag, 2x ecg & 50  \\
% PAMAP2~\cite{pamap2} & 9 & 18 & accel, gyro, mag  & 100  \\
% OPPORTUNITY~\cite{opportunity} & 4 & 3 & accel, gyro, mag & 30  \\
% MotionSense~\cite{motionsense} & 9 & 6 & accel, gyro & 50  \\
% DSADS~\cite{dsads} & 8 & 19 & 5x accel, 5x gyro, 5x mag & 25  \\
% Daphnet~\cite{daphnet} & 10 & 2 & 3x accel & 64  \\
% KU-HAR~\cite{ku-har} & 90 & 18 & accel, gyro & 100  \\
% HARSense~\cite{harsense} & 12 & 6 & accel, gyro & 25  \\ \hline
% \end{tabular}
% \label{tab:whar-datasets}
% \end{table*}
\section{Experiments}

% This section details the experimental setup, including model training and preprocessing performance analysis.

\subsection{Model Training and Evaluation}

\begin{figure*}[h!]
\centering
\includegraphics[width=0.9\linewidth]{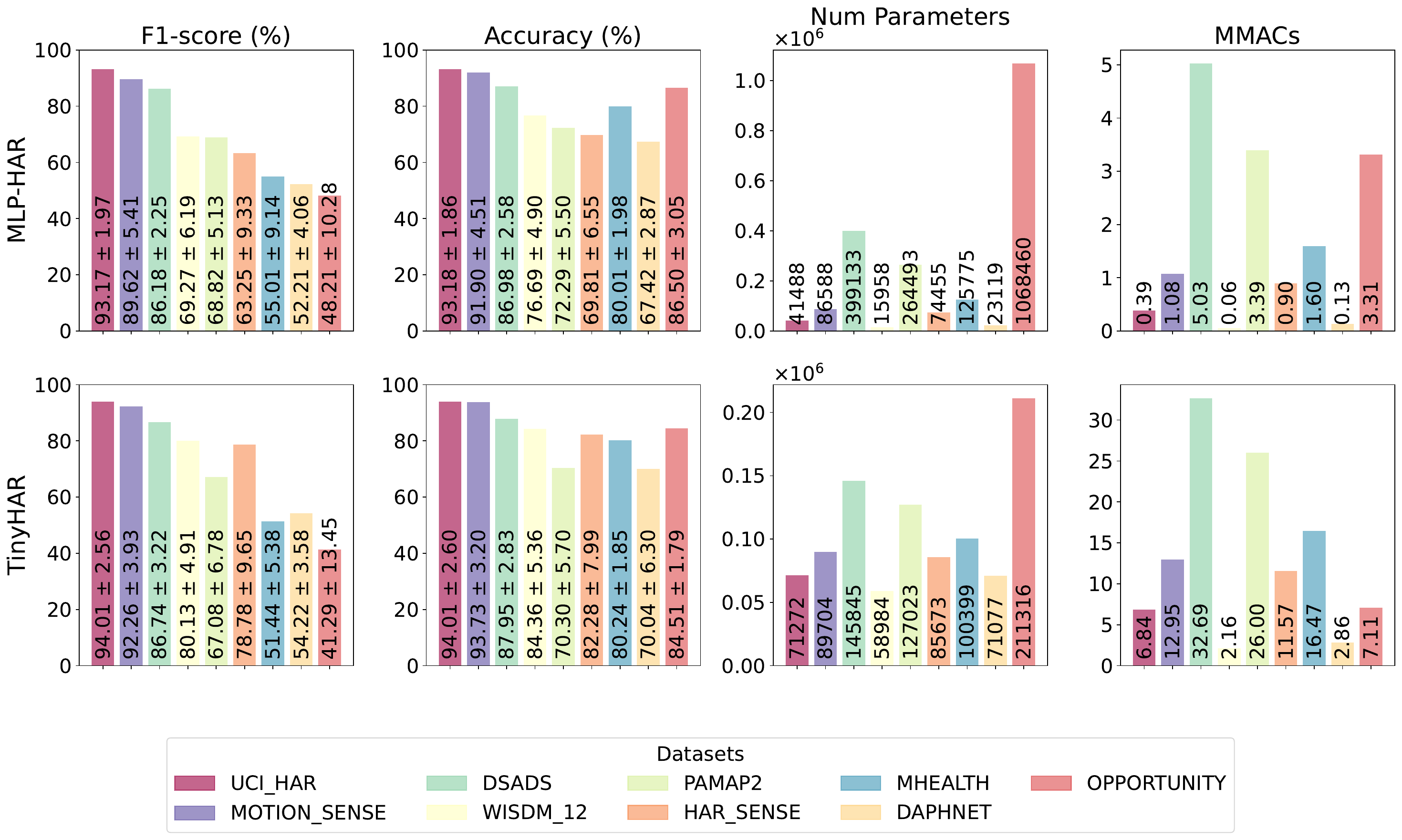}
\caption{Experimental results evaluated on all ten natively supported datasets for two HAR-models: TinyHAR \cite{tinyhar} and MLP-HAR \cite{mlphar}. Reported metrics include Accuracy, Macro F1-Score, the number of parameters and the Multiply-Accumulate (MAC) operations in the millions. The reported results are the mean results of the Leave-One-Subject-Out cross-validation strategy.}
\label{fig:training_results}
\Description{This figure presents a multi-panel bar chart comparing the performance of two models, TinyHAR and MLP-HAR, across different human activity recognition datasets based on F1-score, Accuracy, Number of Parameters, and MMACS (Multiply-Accumulate Operations).
The top row of charts shows the F1-score and Accuracy percentages.
The left chart in the top row, labeled "F1-score (\%)", shows performance for TinyHAR (light blue bars) and MLP-HAR (orange bars) on datasets including UCI_HAR, WISDM_12, PAMAP2, MOTION_SENSE, OPPORTUNITY, MHEALTH, DSADS, DAPHNET, and HAR_SENSE. TinyHAR consistently achieves high F1-scores, for instance, 94.01\% 
± 2.56 for UCI_HAR and 93.17\% 
± 1.97 for WISDM_12. MLP-HAR generally has lower F1-scores, such as 86.74% 
± 3.22 for UCI_HAR and 86.18\% 
± 2.25 for WISDM_12.
The right chart in the top row, labeled "Accuracy (\%)", follows the same structure, comparing TinyHAR (light blue) and MLP-HAR (orange) accuracies. TinyHAR shows accuracies like 94.01\% 
± 2.60 for UCI_HAR and 93.18\% 
± 1.86 for WISDM_12. MLP-HAR's accuracies include 91.90\% 
± 4.51 for UCI_HAR and 87.95\% 
± 2.83 for WISDM_12.
The bottom row of charts focuses on model complexity, showing the Number of Parameters and MMACS.
The left chart in the bottom row, labeled "Num Parameters", displays the number of parameters for TinyHAR (light blue) and MLP-HAR (orange), with values in the thousands. TinyHAR models generally have significantly fewer parameters, such as 41,488 for UCI_HAR , compared to MLP-HAR, which has 86,588 parameters for UCI_HAR. For the MOTION_SENSE dataset, TinyHAR has 15,958 parameters , while MLP-HAR has 264,493 parameters.
The right chart in the bottom row, labeled "MMACS", presents the Multiply-Accumulate Operations. TinyHAR models consistently show much lower MMACS values. For example, UCI_HAR has 6.84 MMACS for TinyHAR and 0.39 MMACS for MLP-HAR. For MOTION_SENSE, TinyHAR has 1.08 MMACS , while MLP-HAR has 32.69 MMACS. Notably, the MMACS scale for MLP-HAR is considerably higher than for TinyHAR across all datasets shown.
The x-axis for all charts lists the datasets: UCI_HAR, WISDM_12, PAMAP2, MOTION_SENSE, OPPORTUNITY, MHEALTH, DSADS, DAPHNET, and HAR_SENSE.}
\end{figure*}

To implicitly demonstrate the functionality of the library, its dataset support, and its usability for benchmarking, we trained two popular WHAR models, TinyHAR \cite{tinyhar} and MLP-HAR \cite{mlphar}, on the 9 natively supported datasets (see \autoref{tab:whar-datasets}). Results are shown in \autoref{fig:training_results}. We did not perform hyperparameter tuning or other performance optimizations such as data cleaning, as the goal was simply to illustrate the usability and ease of benchmarking with the library. As a result, the reported performance is not directly comparable to results from papers that involve model-specific tuning or preprocess the dataset differently. Due to the library's design, the training and evaluation process only requires to implement a new model together with a training script as all the necessary components to obtain the dataloader are included in the library, therefore making this a fair comparison between the two tested models. Training was performed on a single NVIDIA A100 GPU with 40GB of memory, using a batch size of 256, the Adam optimizer, and a learning rate of 0.001. To prevent overfitting, early stopping was implemented, halting training after 15 consecutive epochs without improvement in validation loss. Evaluation followed a Leave-One-Subject-Out (LOSO) cross-validation protocol, with test subjects for each split predefined in the dataloader to ensure consistent and reproducible splits in line with the library’s framework.

\subsection{Preprocessing Performance Analysis}

\begin{figure*}[h!]
\centering
\includegraphics[width=0.9\linewidth]{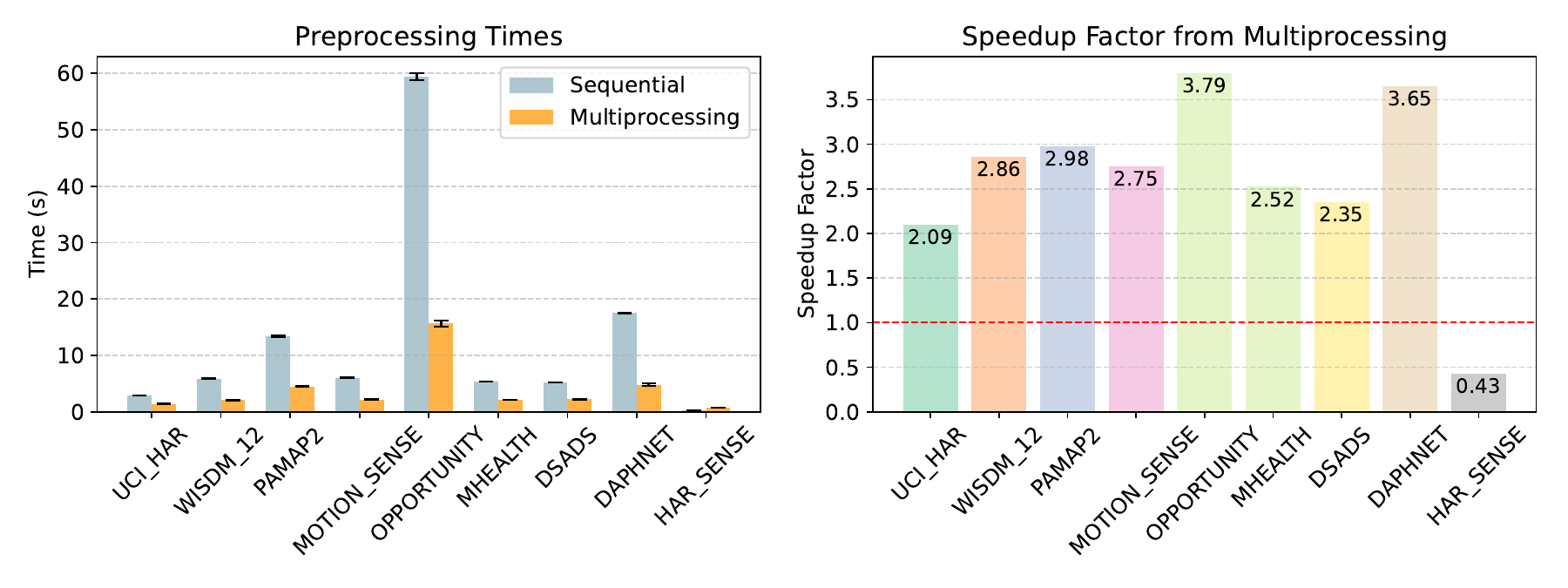}
\caption{Preprocessing performance across the 9 natively supported datasets. The left plot displays the absolute time differences between sequential and multiprocessing execution, while the right plot shows the corresponding speedup factors. Speedup values above 1 (indicated by the red line) represent performance gains from multiprocessing. All measurements were conducted on an M2 MacBook Pro with 10 CPU cores.}
\label{fig:times}
\Description{This figure consists of two bar charts comparing preprocessing times and the resulting speedup factors when using sequential versus multiprocessing approaches for preprocessing various WHAR datasets.
The top chart, titled "Preprocessing Times", displays the time in seconds on the y-axis, ranging from 0 to 60. The x-axis lists nine different datasets: UCI\_HAR, WISDM\_12, PAMAP2, MOTION\_SENSE, OPPORTUNITY, MHEALTH, DSADS, DAPHNET, and HAR\_SENSE. Each dataset is represented by two bars: a light blue bar indicating "Sequential" processing time and an orange bar indicating "Multiprocessing" time. For most datasets, multiprocessing significantly reduces the preprocessing time. For instance, MOTION\_SENSE shows a sequential time nearing 60 seconds, which is reduced to approximately 16 seconds with multiprocessing. DAPHNET also shows a considerable reduction from around 18 seconds sequentially to about 5 seconds with multiprocessing.
The bottom chart, titled "Speedup Factor from Multiprocessing", illustrates the efficiency gain. The y-axis represents the speedup factor, ranging from 0.0 to over 3.5. A red dashed line at 1.0 on the y-axis signifies no change in performance. The x-axis lists the same nine datasets as the top chart. Most datasets exhibit a speedup factor greater than 1.0, indicating faster preprocessing with multiprocessing. MOTION\_SENSE has the highest speedup factor at 3.79, followed closely by DAPHNET at 3.65. Other notable speedups include PAMAP2 at 2.98, WISDM\_12 at 2.86, OPPORTUNITY at 2.75, MHEALTH at 2.52, DSADS at 2.35, and UCI\_HAR at 2.09. Conversely, HAR\_SENSE shows a speedup factor of 0.43, indicating that multiprocessing was slower for this particular dataset.}
\end{figure*}

Furthermore, we assessed the impact of multiprocessing on preprocessing performance in comparison to sequential execution across the 9 natively supported datasets. As illustrated in Figure \ref{fig:times}, we report both absolute time differences and speedup factors, based on measurements conducted on an M2 MacBook Pro with 10 CPU cores. For 8 out of the 9 datasets, multiprocessing achieved speedups between approximately 2.1× and 3.8×. The only exception was HARSense, the smallest dataset, where the overhead of multiprocessing outweighed its benefits. These findings demonstrate notable time savings when preprocessing multiple datasets, underscoring the library’s computational efficiency and scalability to larger datasets.

\section{Conclusion and Future Work}

This paper introduces WHAR Datasets, an open-source library designed to standardize and streamline data handling for Wearable Human Activity Recognition (WHAR). By addressing challenges related to inconsistencies in data structures and formats, which often require dataset-specific handling, the library aims to improve research efficiency, enhance reproducibility, and enable fairer comparability within the WHAR community.

Key contributions of the library include a standardized data format, a configuration- and parser-based approach for dataset integration, WHAR-specific preprocessing and other functional features built on this format, and framework-agnostic adapters compatible with popular deep learning frameworks. These features allow researches to concentrate on application development rather than data handling, facilitating faster experimentation and more reliable comparison across different approaches. The design emphasizes extensibility, enabling straightforward integration of new datasets and ensuring the library’s long-term applicability.

Currently, the library includes a curated collection of 9 widely used WHAR datasets. Future efforts focus on expanding the repository by adding more datasets from the literature, along with advanced data cleaning, augmentation and preprocessing techniques. New functional features are planned, such as window-level auxiliary feature generation (e.g., spectrograms, statistical summaries) alongside raw sensor data, enabling hybrid modeling approaches. We also aim to encourage community contributions to further enrich the library and support the development of standardized benchmarks for WHAR research.

%%
%% The acknowledgments section is defined using the "acks" environment
%% (and NOT an unnumbered section). This ensures the proper
%% identification of the section in the article metadata, and the
%% consistent spelling of the heading.
\begin{acks}
This work is supported by the Helmholtz Association Initiative and Networking Fund on the HAICORE@KIT partition.
\end{acks}

%%
%% The next two lines define the bibliography style to be used, and
%% the bibliography file.
\bibliographystyle{ACM-Reference-Format}
\bibliography{bibliography}

%%
%% If your work has an appendix, this is the place to put it.
%\appendix

\end{document}